\documentclass[12pt]{article}

\def\be{\begin{equation}}
\def\ee{\end{equation}}

\newtheorem{theorem}{Theorem}
\newtheorem{lemma}[theorem]{Lemma}

\newtheorem{definition}[theorem]{Definition}

\begin{document}

\title{Replica symmetry breaking \\ related to a general ultrametric space III: \\
the case of general measure}

\author{A.Yu.Khrennikov\footnote{International Center for Mathematical
Modelling in Physics and Cognitive Sciences, University of
V\"axj\"o, S-35195, Sweden, e--mail:
Andrei.Khrennikov@msi.vxu.se}, S.V.Kozyrev\footnote{Steklov
Mathematical Institute, Moscow, Russia, e--mail:
kozyrev@mi.ras.ru}}

\maketitle

\begin{abstract}
Family of replica matrices, related to general ultrametric spaces
with general measures, is introduced. These matrices generalize
the known Parisi matrices. Some functionals of replica approach
are computed. Replica symmetry breaking solution is found.
\end{abstract}

\section{Introduction}

In the present paper we introduce a very general replica symmetry
anzats, and, using this anzats, find a family of replica symmetry
breaking solutions with functional freedom. We continue to develop
(see also \cite{Prep1}, \cite{Prep2}) the new replica symmetry
breaking (RSB) anzats, related to pseudodifferential operators on
general ultrametric spaces. We consider the case, in which the
measure $\nu$ at the ultrametric space $X$ under consideration may
be arbitrary. We introduce replica matrix of the form, see the
next Section for the notations, \be\label{newmatrix}
Q_{IJ}=\sqrt{\nu(I)\nu(J)}T{(\,{\rm sup}\,(I,J))},\qquad I,J\in
{\cal S}_{\rm min} \ee which we call {\it the {\rm sup}--matrix}.
This matrix is defined by the function $T{(I)}$ on the finite
subtree ${\cal S}$ of some (probably, infinite) directed tree
${\cal T}(X)$, related to the ultrametric space $X$, and by
arbitrary measure $\nu$ on $X$ (with the only restriction that
measures of all balls in $X$ are positive). The difference of this
matrix from the considered in \cite{Prep1}, \cite{Prep2} is that
in the present paper the measure $\nu$ is arbitrary.

This approach is motivated by the results of \cite{ABK},
\cite{PaSu}, where the equivalence between the family of the
Parisi matrices and $p$--adic pseudodifferential operators was
shown, in the sense that the Parisi matrix allows some
renumeration of rows and columns, that makes the matrix element
dependent only on the $p$--adic norm of the difference between its
indices:
$$
Q_{ab}=q(|a-b|_p)
$$
and, therefore, the corresponding Parisi matrix can be
diagonalized by $p$--adic Fourier transform. Discussion of more
general Parisi matrices and the corresponding Fourier transforms
may be found in \cite{Carlucci1}. This result shows that replica
symmetry breaking is related to theory of $p$--adic PDO
(pseudodifferential operators), such as the Vladimirov operator
$D^{\alpha}$ \cite{VVZ}. For other developments in $p$--adic
mathematical physics see \cite{wavelets}--\cite{ABKO}. For results
of replica method see \cite{MPV}.

In the present paper we, using the theory of PDO on general
ultrametric spaces \cite{ACHA}, \cite{Izv}, \cite{nextIzv}, find a
very general form (\ref{newmatrix}) of replica matrix, which is
obtained by restricting of ultrametric PDO on finite dimensional
space of test functions.

We define the variant of the $n\to0$ limit procedure for the
replica anzats under consideration and a variant of analysis on
trees. Varying the free energy in the vicinity of phase
transition, we find the replica symmetry breaking equation, which
is some integro--differential equation on the tree ${\cal S}$ in
the sense of the introduced tree analysis (i.e. it has the form of
equation with summation and taking of finite differences on the
tree). In the case under consideration we are able to find only
one solution of the RSB equation --- the constant solution, for
which $T{(J)}=T={\rm const}$ (we do not claim that there is no any
other solution, but we are not able to find these solutions
analytically). Note that in the case under consideration the
constant solution (contrary to the case of the Parisi anzats) is
highly nontrivial, since in this case the replica matrix depends
on arbitrary measure $\nu$:
$$
Q_{IJ}=T\sqrt{\nu(I)\nu(J)}
$$
and $T$ is some constant. This means that we found the family of
RSB solutions with a functional freedom.

Most of the results, presented in this paper, was obtained in
papers \cite{Prep1}, \cite{Prep2} of the authors for more
particular choice of replica matrix, which is obtained from
(\ref{newmatrix}) by fixing the measure $\nu$ in such a way that
the measures of maximal subbals in any ball will be equal.
Moreover, for this particular case also some analog of the Parisi
replica symmetry breaking solution was obtained.

The structure of the present paper is as follows.

In Section 2 we develop the form of the replica sup--matrix, using
the ultrametric pseudodifferential sup--operator (see the Appendix
for discussion) of \cite{ACHA}, \cite{Izv}, \cite{nextIzv}.

In Section 3 we develop analysis on trees, see also our previous
preprints \cite{Prep1}, \cite{Prep2}, for the considered case of
general measure.

In Section 4 we compute some simple sums for the introduced
replica matrices.

In Sections 5 and 6 we generalize the definitions and computations
of \cite{Prep1}, \cite{Prep2} of the $n\to0$ limit procedure for
general ultrametric spaces onto the case under consideration.

In Section 7 we find the replica symmetry breaking equation.

In Section 8 we find the constant solution of the replica symmetry
breaking equation.

In Section 9 (the Appendix) we put some materials on ultrametric
pseudodifferential operators from \cite{nextIzv}.

\section{Ultrametric PDO and replica matrices}

In the present section we show that the mentioned block matrices
(\ref{newmatrix}) are related to ultrametric pseudodifferential
operators (PDO) on ultrametric spaces, considered in \cite{Izv},
\cite{nextIzv}, see the Appendix for discussion.Our construction
is based on a directed tree, which we will denote ${\cal T}$ (and
which is a tree of balls ${\cal T}(X)$ in some ultrametric space
$X$). Consider a subtree ${\cal S}\subset {\cal T}$ satisfying the
following definition.

\begin{definition}\label{wave_type}
{\sl The subset ${\cal S}$ in a directed tree ${\cal T}$ (with the
partial order of the kind considered in the Appendix) is called of
the regular type, iff:

1) ${\cal S}$ is finite;

2) ${\cal S}$ is a directed subtree in ${\cal T}$ (where the
direction in ${\cal S}$ is the restriction of the direction in
${\cal T}$ onto ${\cal S}$);

3) The directed subtree ${\cal S}$ obey the following property: if
${\cal S}$ contains a vertex $I$ and a vertex $J$: $J<I$,
$|IJ|=1$, then the subtree ${\cal S}$ contains all the vertices
$L$ in ${\cal T}$: $L<I$, $|IL|=1$. }
\end{definition}

Here $|IJ|$ is the distance between vertices of the tree (number
of links in the path connecting these vertices).

The maximal vertex in ${\cal S}$ we will denote $K$.

We consider the ultrametric space $X({\cal T})$ with arbitrary
measure $\nu$ (such that the measure of any ball is positive).
Vertices of the tree ${\cal T}={\cal T}(X)$ are in one to one
correspondence with balls (or disks) in $X$. We denote by $J$ the
ball in $X$, corresponding to vertex $J\in {\cal T}(X)$, and by
$\chi_J$ we denote the characteristic function of this ball.

For the finite subtree ${\cal S}\subset {\cal T}$ of the regular
type consider the space $D({\cal S})$ \footnote{The space $D({\cal
S})$ is a finite dimensional space of test functions on the
ultrametric space $X$ and is an analogue of the Bruhat--Schwartz
space of test functions of $p$--adic argument. The space $D(X)$ of
test functions on $X$ is the inductive limit of the spaces
$D({\cal S})$: $D(X)=\lim\,{\rm ind}_{{\cal S}\to {\cal T}}\,
D({\cal S})$.}, which is the linear span of vectors $\chi_J$ with
$J\in {\cal S}$. We consider this space as the subspace in the
space $L^2(X,\nu)$ of quadratically integrable with respect to the
measure $\nu$ functions on $X$. Obviously in the space $D({\cal
S})$ there is the orthonormal basis $\left\{{\chi_J\over
\sqrt{\nu(J)}}\right\}$ with $J$ running over the minimal vertices
in ${\cal S}$, i.e. a vector $f$ in $D({\cal S})$ can be put into
the form \be\label{funinS} f=\sum_{J\in {\cal S}_{\rm min}}f_J
{\chi_J\over \sqrt{\nu(J)}} \ee where the summation runs over
minimal elements in ${\cal S}$.

Consider the operator $Q({\cal S})=\Pi({\cal S})T$ in $D({\cal
S})$, where $\Pi({\cal S})$ is the orthogonal projection in the
space $L^2(X,\nu)$ onto $D({\cal S})$, and $T$ is an ultrametric
pseudodifferential operator (PDO), see the Appendix. Operator
$Q({\cal S})$ is an operator in finite dimensional linear space
$D({\cal S})$.

\begin{lemma}\label{theformula}{\sl
Action of $Q({\cal S})=\Pi({\cal S})T$ on the functions
(\ref{funinS}) in $D({\cal S})$ is given by the formula
$$
(Q({\cal S})f)_I=\sum_{J\in {\cal S}_{\rm
min}}\sqrt{\nu(I)\nu(J)}T{(\,{\rm
sup}\,(I,J))}\left[\sqrt{\nu(J)\over\nu(I)}f_I -
f_J\right]+C_{\cal S}f_I;
$$
where
$$
C_{\cal S}=\int_{y\notin K }
d\nu(y)T(I,y)=\sum_{L>K}T{(L)}\left(\nu(L)-\nu({L-1,K})\right)
$$
and $I$, $J$ run over the minimal vertices in ${\cal S}$. Here
${L-1,K}$ is the maximal subball in $L$, which is larger than
$K$.}
\end{lemma}

\noindent{\it Proof}\qquad Take $I\in {\cal S}_{\rm min}$ and
consider the scalar product in $L^2(X,\nu)$:
$$
\left\langle {\chi_I\over \sqrt{\nu(I)}}, T\sum_{J\in {\cal
S}_{\rm min}}f_J {\chi_J\over \sqrt{\nu(J)}}\right\rangle=$$
$$
=\sum_{J\in {\cal S}_{\rm min}}{1\over
\sqrt{\nu(I)\nu(J)}}f_J\int_{x\in I}d\nu(x)\int
d\nu(y)T(x,y)(\chi_J(x)-\chi_J(y))=
$$
$$
={1\over \nu(I)}f_I\int_{x\in I}d\nu(x)\int
d\nu(y)T(x,y)(\chi_I(x)-\chi_I(y))+
$$
$$
+\sum_{J\in {\cal S}_{\rm min},J\ne I}{1\over
\sqrt{\nu(I)\nu(J)}}f_J\int_{x\in I}d\nu(x)\int
d\nu(y)T(x,y)(\chi_J(x)-\chi_J(y))=
$$
$$
={1\over \nu(I)}f_I\int_{x\in I}d\nu(x)\int_{y\notin I }
d\nu(y)T(x,y)-
$$
$$
-\sum_{J\in {\cal S}_{\rm min},J\ne I}{1\over
\sqrt{\nu(I)\nu(J)}}T{(\,{\rm sup}\,(I,J))}f_J\int_{x\in
I}d\nu(x)\int d\nu(y)\chi_J(y)=
$$
$$
=f_I\int_{y\notin I } d\nu(y)T(I,y)- \sum_{J\in {\cal S}_{\rm
min},J\ne I}\sqrt{\nu(I)\nu(J)}T{(\,{\rm sup}\,(I,J))}f_J=
$$
$$
=f_I\int_{y\notin K } d\nu(y)T(I,y)+f_I\sum_{J\in {\cal S}_{\rm
min},J\ne I}\nu(J)T{(\,{\rm sup}\,(I,J))}-
$$
$$
-\sum_{J\in {\cal S}_{\rm min},J\ne I}\sqrt{\nu(I)\nu(J)}T{(\,{\rm
sup}\,(I,J))}f_J=
$$
$$
=f_I\int_{y\notin K } d\nu(y)T(I,y)+\sum_{J\in {\cal S}_{\rm
min}}\sqrt{\nu(I)\nu(J)}T{(\,{\rm
sup}\,(I,J))}\left[\sqrt{\nu(J)\over\nu(I)}f_I - f_J\right].
$$
This proves the lemma.

\bigskip

Therefore, the operator $Q({\cal S})$ corresponds to the block
matrix: \be\label{Q} Q_{IJ}=\sqrt{\nu(I)\nu(J)}T{(\,{\rm
sup}\,(I,J))},\qquad I,J\in {\cal S}_{\rm min} \ee This kind of
replica matrix we call {\it the {\rm sup}--matrix}.

\bigskip

\noindent{\bf Remark}\qquad Note that $T(I,y)$ in the expression
for $C_{\cal S}$ does not depend on $I\in {\cal S}$ and depends
only on $K={\rm sup}\, {\cal S}$. The constant $C_{\cal S}$ tends
to zero if the largest vertex of ${\cal S}$ tends to infinity. We
will ignore the contribution containing $C_{\cal S}$ in the
following.

\section{Analysis on trees}

In the present section we discuss the analysis on directed trees
${\cal T}(X)$ of balls in $X$. We define tree derivation and
integration over the subtrees of regular type. These results
generalize the results of \cite{Prep1}, \cite{Prep2}.

\begin{definition}{\sl For the function $F(J)$ on the directed tree ${\cal
T}$ the function \be\label{Delta} \Delta F(J)=
F(J)-\sum_{j=0,\dots,p_J-1:J_j<J,|JJ_j|=1}F(J_j) \ee we call the
tree derivative.

The tree integral over the subtree of the regular type ${\cal
S}\subset {\cal T}$ we define as
$$
\int_{\cal S} F=\sum_{J\in {\cal S}\backslash {\cal S}_{\rm
min}}F(J)
$$
}
\end{definition}
Here ${\cal S}_{\rm min}$ is the set of minimal elements in ${\cal
S}$, distance $|IJ|$ between vertices of the tree is the number of
edges in the path connecting $I$ and $J$, $p_J$ is the branching
index of vertex $J$ in the tree ${\cal T}$. The branching index of
vertex $J$ is equal to $p_J$ if $J$ is incident to $p_J$
decreasing links in ${\cal T}$ with the beginning in $J$. In the
language of the corresponding ultrametric space $X$ vertex $J$ has
the branching index $p_J$, if the corresponding ball $J$ has $p_J$
maximal subballs.

In the following in the notation for the tree derivative for
simplicity instead of (\ref{Delta}) we use the simplified notation
$$
\Delta F(J)= F(J)-\sum_{j}F(J_j)
$$

The next lemma relates the analysis on the directed tree and the
analysis on the corresponding ultrametric space. This shows, that
generalized functions at $X$ can be considered as the constants of
tree derivation at ${\cal T}$.

\begin{lemma}
{\sl The space of solutions of the equation
$$
\Delta F(J)=0
$$
is isomorphic, as a linear space, to the space of generalized
functions at $X$, with the isomorphism defined by the formula
$$
\phi_F(\chi_J)=F(J)
$$
Here $\phi_F$ is the generalized function at $X$, corresponding to
the function $F$ at the tree, and $\chi_J$ is the characteristic
function of the disk $J$. }
\end{lemma}

\noindent{\it Proof}\qquad The proof is by the remark that a
generalized function on the ultrametric space $X$ is defined
unambiguously by its values (as of the functional) on the
characteristic functions of disks.

Characteristic functions of disks are not linearly independent,
but are related as follows
$$
\chi_{J}=\sum_{j=0,\dots,p_J-1:J_j<J,|JJ_j|=1}\chi_{J_j}
$$
By linearity of generalized functions, this implies the following
conditions of the values of generalized functions:
$$
\phi(\chi_J)=\sum_{j=0,\dots,p_J-1:J_j<J,|JJ_j|=1}\phi(\chi_{J_j})
$$
Choosing $F(J)=\phi(\chi_J)$, we get exactly
$$
\Delta F(J)=0
$$
Since no other restrictions on $F(J)$ are put, this proves the
lemma.

\bigskip

There exist several analogies between the introduced analysis on
trees and the analysis of functions of real argument. We have the
following partial analogue of the Leibnitz rule
\be\label{LeibRule} \Delta
F(J)G(J)=F(J)G(J)-\sum_{j=0,\dots,p_J-1:J_j<J,|JJ_j|=1}F(J_j)G(J_j)=
$$
$$
=F(J)G(J)-\sum_{j}F(J_j)G(J)+\sum_{j}F(J_j)G(J)-\sum_{j}F(J_j)G(J_j)=
$$
$$
=[F(J)-\sum_{j}F(J_j)]G(J)+\sum_{j}F(J_j)[G(J)-G(J_j)]=
$$
$$
=[F(J)-\sum_{j}F(J_j)]G(J)+F(J)\sum_{j}[G(J)-G(J_j)]-\sum_{j}[F(J)-F(J_j)][G(J)-G(J_j)]
\ee

The next lemma gives the tree analogue of the Newton--Leibnitz
formula
$$
\int_a^b{df(x)\over dx}dx=f(b)-f(a)
$$
This lemma shows the importance of the notion of a subtree of the
regular type.

\begin{lemma}{\sl For the subtree ${\cal S}\subset {\cal T}$ of the regular type the following tree Newton--Leibnitz
formula is satisfied
$$
\int_{{\cal S}}\Delta F(J)=F(K)-\sum_{J\in
 {\cal S}_{\rm min}}F(J)
$$

}
\end{lemma}

In the next two formulas the tree derivative is taken with respect
to $J$.

\begin{lemma}{\sl
\be\label{higher} \Delta \sum_{L\in {\cal S}\backslash {\cal
S}_{\rm min}:L\le J}F(L)=F(J) \ee \be\label{lower} \Delta
\left[\nu(J)\sum_{L:J\le L\le
K}F(L)\right]=-\sum_{j=0,\dots,p_J-1:J_j<J,|JJ_j|=1}\nu({J_j})F(J_j)
\ee }
\end{lemma}

The formula (\ref{higher}) (respectively (\ref{lower})) is the
tree analogue of the derivative of the integral over the higher
(respectively the lower) limit.

\bigskip

The next lemma is the analogue of the following change of order of
integration:
$$
\int_a^b
f(x)\left[\int_a^xg(y)dy\right]dx=\int_a^bg(y)\left[\int_y^bf(x)dx\right]dy
$$

\begin{lemma}\label{change}{\sl
$$
\sum_{L\in {\cal S}\backslash {\cal S}_{\rm min}} F(L)\sum_{B\in
{\cal S}\backslash {\cal S}_{\rm min} :B<L}G(B)=\sum_{L\in {\cal
S}\backslash {\cal S}_{\rm min}}G(L)\sum_{B:L<B\le K}F(B)
$$
}
\end{lemma}

The above sums are the analogues of integration over several
variables.

\section{Computations with replica matrices}

In the present section we compute the products and traces of the
degrees for replica matrices under investigation. Some similar
computations for $p$--adic case can be found in \cite{PaSu}. The
next lemma gives some useful formulas for computations on the
subtrees of the regular type.

\begin{lemma}\label{formulae}
{\sl
$$
\sum_{J\in {\cal S}_{\rm min}:\,{\rm
sup}\,(I,J)=L}\nu(J)=\nu(L)-\nu({L-1,I})
$$
$$
\sum_{I,J\in {\cal S}_{\rm min}:\,{\rm
sup}\,(I,J)=L}\nu(I)\nu(J)=\nu^2(L)-\sum_{j:L_j<L,|LL_j|=1}\nu^2({L_j})=\Delta
\nu^2(L)
$$
}
\end{lemma}

Here ${L-1,I}$ is the maximal subball in $L$, which contains the
ball $I$.

Proof of the lemma is by additivity of the measure $\nu$.

The following lemma computes the important in the replica approach
functional $\sum_{a\ne b}Q_{ab}$.

\begin{lemma}\label{traceRMst}
{\sl
$$
\sum_{I,J\in {\cal S}_{\rm min};I\ne J}\nu(I){\nu(J)}{T{(\,{\rm
sup}\,(I,J))}}= \sum_{J\in {\cal S}\backslash {\cal S}_{\rm
min}}{T{(J)}}\Delta\nu^2(J)
$$
}
\end{lemma}

\noindent{\it Proof}\qquad Computing the sum over $J$ we get
$$
\sum_{I,J\in {\cal S}_{\rm min};I\ne J}\nu(I){\nu(J)}{T{(\,{\rm
sup}\,(I,J))}}= \sum_{I\in {\cal S}_{\rm min}}\nu(I)\sum_{I< L\le
K}{T{(L)}}\sum_{J\in {\cal S}_{\rm min}:\,{\rm
sup}\,(I,J)=L}{\nu(J)}
$$
where $K$ is the largest vertex in ${\cal S}$. This reduces to
$$
\sum_{I\in {\cal S}_{\rm min}}\nu(I)\sum_{I< L\le
K}{T{(L)}}(\nu(L)-\nu({L-1,I}))= \sum_{J\in {\cal S}\backslash
{\cal S}_{\rm min}}{T{(J)}}\Delta\nu^2(J)
$$
This finishes the proof of the lemma.

\bigskip

Prove the following important lemma, which allows to compute
products of replica matrices in terms of summation over the paths
in corresponding trees.

\begin{lemma}\label{prod_mat}{\sl
For $I,J\in {\cal S}_{\rm min}$ one has: if $I\ne J$, then
$$
\sum_{L\in {\cal S}_{\rm min},L\ne I,J}\nu(L)T_1{(\,{\rm
sup}\,(I,L))}T_2{(\,{\rm sup}\,(L,J))} =
$$
$$
=\sum_{L:\,{\rm sup}\,(I,J)<L\le K}\left(\nu(L)-\nu({L-1,{\rm
sup}(I,J)})\right)T_1{(L)}T_2{(L)}+
$$
$$
+\sum_{L:I<L<\,{\rm
sup}\,(I,J)}\left(\nu(L)-\nu({L-1,I})\right)T_1{(L)}T_2{(\,{\rm
sup}\,(I,J))}+
$$
$$
+\sum_{L:J<L<\,{\rm
sup}\,(I,J)}\left(\nu(L)-\nu({L-1,J})\right)T_1{(\,{\rm
sup}\,(I,J))}T_2{(L)}+
$$
$$
+ \left(\nu\left({\,{\rm sup}\,(I,J)}\right)-\nu\left({\,{\rm
sup}\,(I,J)-1,I}\right)-\nu\left({\,{\rm
sup}\,(I,J)-1,J}\right)\right)T_1{(\,{\rm sup}\,(I,J))}T_2{(\,{\rm
sup}\,(I,J))}
$$
and for $I=J$ one has
$$
\sum_{L\in {\cal S}_{\rm min},L\ne I}\nu(L)T_1{(\,{\rm
sup}\,(I,L))}T_2{(\,{\rm sup}\,(L,I))} =\sum_{L:I<L\le
K}\left(\nu(L)-\nu({L-1,I})\right)T_1{(L)}T_2{(L)}
$$
In particular \be\label{sum_T_IJ} \sum_{L\in {\cal S}_{\rm
min},L\ne I}\nu(L){T{(\,{\rm sup}\,(I,L))}}^2 =\sum_{L:I<L\le
K}\left(\nu(L)-\nu({L-1,I})\right){T{(L)}}^2 \ee

}
\end{lemma}

\noindent{\it Proof}\qquad Compute for $I\ne J$
$$
\sum_{L\in {\cal S}_{\rm min},L\ne I,J}\nu(L)T_1{(\,{\rm
sup}\,(I,L))}T_2{(\,{\rm sup}\,(L,J))} =
$$
$$
=\sum_{L\in {\cal S}_{\rm min},L\ne I,J:\,{\rm sup}\,(L,\,{\rm
sup}\,(I,J))> \,{\rm sup}\,(I,J) }\nu(L)T_1{(\,{\rm
sup}\,(I,L))}T_2{(\,{\rm sup}\,(L,J))}+
$$
$$
+\sum_{L\in {\cal S}_{\rm min},L\ne I,J:\,{\rm sup}\,(L,I)<\,{\rm
sup}\,(I,J)}\nu(L)T_1{(\,{\rm sup}\,(I,L))}T_2{(\,{\rm
sup}\,(L,J))}+
$$
$$
+\sum_{L\in {\cal S}_{\rm min},L\ne I,J:\,{\rm sup}\,(L,J)<\,{\rm
sup}\,(I,J)}\nu(L)T_1{(\,{\rm sup}\,(I,L))}T_2{(\,{\rm
sup}\,(L,J))}+
$$
$$
+\sum_{L\in {\cal S}_{\rm min},L\ne I,J:\,{\rm sup}\,(L,I)=\,{\rm
sup}\,(L,J)=\,{\rm sup}\,(I,J)}\nu(L)T_1{(\,{\rm
sup}\,(I,L))}T_2{(\,{\rm sup}\,(L,J))}
$$
$$
=\sum_{L:\,{\rm sup}\,(I,J)<L\le K}\left(\nu(L)-\nu({L-1,{\rm
sup}(I,J)})\right)T_1{(L)}T_2{(L)}+
$$
$$
+\sum_{L:I<L<\,{\rm
sup}\,(I,J)}\left(\nu(L)-\nu({L-1,I})\right)T_1{(L)}T_2{(\,{\rm
sup}\,(I,J))}+
$$
$$
+\sum_{L:J<L<\,{\rm
sup}\,(I,J)}\left(\nu(L)-\nu({L-1,J})\right)T_1{(\,{\rm
sup}\,(I,J))}T_2{(L)}+
$$
$$
+ \left(\nu\left({\,{\rm sup}\,(I,J)}\right)-\nu\left({\,{\rm
sup}\,(I,J)-1,I}\right)-\nu\left({\,{\rm
sup}\,(I,J)-1,J}\right)\right)T_1{(\,{\rm sup}\,(I,J))}T_2{(\,{\rm
sup}\,(I,J))}
$$
Analogously, for $I=J$ one obtains
$$
\sum_{L\in {\cal S}_{\rm min},L\ne I}\nu(L)T_1{(\,{\rm
sup}\,(I,L))}T_2{(\,{\rm sup}\,(L,I))} =\sum_{L:I<L\le
K}\left(\nu(L)-\nu({L-1,I})\right)T_1{(L)}T_2{(L)}
$$
This finishes the proof of the lemma.

\bigskip

The next lemma gives the trace of the cubic combination of the
replica matrices.

\begin{lemma}\label{trQ3}{\sl For the trace of the cubic combination of the
replica block matrices we get \be\label{ultra1}\,{\rm
tr}\,Q_1Q_2Q_3= \sum_{A,B,C\in {\cal S}_{\rm min},A\ne B,B\ne
C,C\ne A}\nu(A)\nu(B)\nu(C)T_1{(\,{\rm sup}\,(A,B))}T_2{(\,{\rm
sup}\,(B,C))}T_3{(\,{\rm sup}\,(C,A))} =
$$
$$
=\sum_{B\in {\cal S}\backslash {\cal S}_{\rm
min}}T_1{(B)}T_2{(B)}T_3{(B)}\left(3\nu(B)\Delta \nu^2(B)-2\Delta
\nu^3(B)\right)+
$$
$$
+\sum_{B\in {\cal S}\backslash {\cal S}_{\rm
min}}\Delta\nu^2(B)\sum_{L:B<L\le K}(\nu(L)-\nu({L-1,B}))
$$
$$
\left[T_1{(L)}T_2{(L)}T_3{(B)}+T_2{(L)}T_3{(L)}T_1{(B)}+T_3{(L)}T_1{(L)}T_2{(B)}\right]
\ee }
\end{lemma}

The property of the sum in (\ref{ultra1}), that two of the three
(or even three of the three) indices $I$ of the coefficients
$T{(I)}$ coincide, and the two coinciding indices are larger than
the third, is related to the fact that in ultrametric space all
triangles are equilateral with equal larger edges. The analogous
observations for the Parisi RSB anzats were made in \cite{nature}.
Note that here (unlike in \cite{nature}) we did not yet introduce
any kind of the $n\to 0$ limit.

\bigskip

\noindent{\it Proof of Lemma \ref{trQ3}}\qquad Compute
$$
\sum_{A,B,C\in {\cal S}_{\rm min},A\ne B,B\ne C,C\ne
A}\nu(A)\nu(B)\nu(C)T_1{(\,{\rm sup}\,(A,B))}T_2{(\,{\rm
sup}\,(B,C))}T_3{(\,{\rm sup}\,(C,A))} =
$$
$$
=\sum_{A,C\in {\cal S}_{\rm min};C\ne A}\nu(A)\nu(C)T_3{(\,{\rm
sup}\,(C,A))}
$$
$$
\Biggl[ \sum_{B:\,{\rm sup}\,(A,C)<B\le
K}\left(\nu(B)-\nu({B-1,{\rm sup}(A,C)})\right)T_1{(B)}T_2{(B)}+
$$
$$
+\left[\sum_{B:A<B<\,{\rm
sup}\,(A,C)}\left(\nu(B)-\nu({B-1,A})\right)T_1{(B)}\right]T_2{(\,{\rm
sup}\,(A,C))}+
$$
$$
+T_1{(\,{\rm sup}\,(A,C))}\left[\sum_{B:C<B<\,{\rm
sup}\,(A,C)}\left(\nu(B)-\nu({B-1,C})\right)T_2{(B)}\right]+
$$
$$
+ \left(\nu\left({\,{\rm sup}\,(A,C)}\right)-\nu\left({\,{\rm
sup}\,(A,C)-1,A}\right)-\nu\left({\,{\rm
sup}\,(A,C)-1,C}\right)\right)T_1{(\,{\rm sup}\,(A,C))}T_2{(\,{\rm
sup}\,(A,C))} \Biggr]
$$

Discuss the second contribution to the above sum, i.e. the
expression
$$
\sum_{A,C\in {\cal S}_{\rm min};C\ne A}\nu(A)\nu(C)T_2{(\,{\rm
sup}\,(A,C))}T_3{(\,{\rm sup}\,(C,A))}\left[\sum_{B:A<B<\,{\rm
sup}\,(A,C)}\left(\nu(B)-\nu({B-1,A})\right)T_1{(B)}\right]
$$
In this expression we put the summation in the form of the
following composition of the three summations
$$
\sum_{A,C\in {\cal S}_{\rm min};C\ne A}=\sum_{C\in {\cal S}_{\rm
min}}\sum_{L:C<L\le K}\sum_{A\in {\cal S}_{\rm min}:\,{\rm
sup}\,(A,C)=L}
$$
This implies
$$
\sum_{C\in {\cal S}_{\rm min}}\sum_{L:C<L\le K}\sum_{A\in {\cal
S}_{\rm min}:\,{\rm sup}\,(A,C)=L}\nu(A)\nu(C)T_2{(\,{\rm
sup}\,(A,C))}T_3{(\,{\rm sup}\,(C,A))}$$
$$
\left[\sum_{B:A<B<\,{\rm
sup}\,(A,C)}\left(\nu(B)-\nu({B-1,A})\right)T_1{(B)}\right]=
$$
$$
=\sum_{C\in {\cal S}_{\rm min}}\nu(C)\sum_{L:C<L\le
K}T_2{(L)}T_3{(L)} \sum_{A\in {\cal S}_{\rm min}:\,{\rm
sup}\,(A,C)=L}\nu(A)\left[\sum_{B:A<B<L}\left(\nu(B)-\nu({B-1,A})\right)T_1{(B)}\right]=
$$
$$
=\sum_{C\in {\cal S}_{\rm min}}\nu(C)\sum_{L:C<L\le
K}T_2{(L)}T_3{(L)} \left[\sum_{B\in {\cal S}\backslash {\cal
S}_{\rm min} :B<L,\,{\rm
sup}\,(B,C)=L}T_1{(B)}\Delta\nu^2(B)\right]=
$$
$$
=\sum_{L\in {\cal S}\backslash {\cal S}_{\rm
min}}T_2{(L)}T_3{(L)}\sum_{C\in {\cal S}_{\rm min}:C<L}\nu(C)
\left[\sum_{B\in {\cal S}\backslash {\cal S}_{\rm min} :B<L,\,{\rm
sup}\,(B,C)=L}T_1{(B)}\Delta\nu^2(B)\right]=
$$
$$
=\sum_{L\in {\cal S}\backslash {\cal S}_{\rm min}}T_2{(L)}T_3{(L)}
\left[\sum_{B\in {\cal S}\backslash {\cal S}_{\rm min}
:B<L}T_1{(B)}(\nu(L)-\nu({L-1,B}))\Delta\nu^2(B)\right]=
$$
\be\label{the3rd} = \sum_{B\in {\cal S}\backslash {\cal S}_{\rm
min}}T_1{(B)}\Delta\nu^2(B)\left[\sum_{L:B<L\le
K}T_2{(L)}T_3{(L)}(\nu(L)-\nu({L-1,B})) \right]\ee

Analogously, the third contribution
$$
\sum_{A,C\in {\cal S}_{\rm min};C\ne A}\nu(A)\nu(C)T_1{(\,{\rm
sup}\,(A,C))}T_3{(\,{\rm sup}\,(C,A))}\left[\sum_{B:C<B<\,{\rm
sup}\,(A,C)}\left(\nu(B)-\nu({B-1,C})\right)T_2{(B)}\right]
$$
will give \be\label{the4rd}  \sum_{B\in {\cal S}\backslash {\cal
S}_{\rm min}}T_2{(B)}\Delta\nu^2(B)\left[\sum_{L:B<L\le
K}T_1{(L)}T_3{(L)}(\nu(L)-\nu({L-1,B})) \right]\ee

Compute the first contribution. We get
$$
\sum_{A,C\in {\cal S}_{\rm min};C\ne A}\nu(A)\nu(C)T_3{(\,{\rm
sup}\,(C,A))} \left[ \sum_{B:\,{\rm sup}\,(A,C)<B\le
K}\left(\nu(B)-\nu({B-1,{\rm
sup}(A,C)})\right)T_1{(B)}T_2{(B)}\right]=
$$
\be\label{the1st}  =\sum_{B\in {\cal S}\backslash {\cal S}_{\rm
min}}T_3{(B)}\Delta\nu^2(B)\left[\sum_{L:B<L\le
K}T_1{(L)}T_2{(L)}(\nu(L)-\nu({L-1,B})) \right]\ee

The fourth contribution reduces to

$$
\sum_{A,C\in {\cal S}_{\rm min};C\ne A}\nu(A)\nu(C)T_1{(\,{\rm
sup}\,(A,C))}T_2{(\,{\rm sup}\,(A,C))}T_3{(\,{\rm sup}\,(C,A))}
$$
$$
\left(\nu\left({\,{\rm sup}\,(A,C)}\right)-\nu\left({\,{\rm
sup}\,(A,C)-1,A}\right)-\nu\left({\,{\rm
sup}\,(A,C)-1,C}\right)\right)=
$$
$$
=\sum_{B\in {\cal S}\backslash {\cal S}_{\rm
min}}T_1{(B)}T_2{(B)}T_3{(B)}\left(\nu(B)\Delta
\nu^2(B)-2\sum_{j:B_j<B,|BB_j|=1}\nu^2({B_j})\left(\nu({B})-\nu({B_j})\right)\right)=
$$
\be\label{the2nd} =\sum_{B\in {\cal S}\backslash {\cal S}_{\rm
min}}T_1{(B)}T_2{(B)}T_3{(B)}\left(3\nu(B)\Delta \nu^2(B)-2\Delta
\nu^3(B)\right) \ee

Combining formulas (\ref{the3rd})--(\ref{the2nd}), we obtain the
proof of the lemma.

\section{The $n\to 0$ limit: definition}

The present and the next sections are written at the physical
level of rigor. In the present section we describe the
generalization of the $n\to 0$ limit of replica approach, relevant
to the introduced replica symmetry breaking (RSB) anzats.

Consider the map $\eta$, which acts on the measures $\nu(J)$ of
ultrametric disks according to the following rules:

1) Normalization: \be\label{norm42} \eta(\nu(R))=1 \ee where $R$
is the root of the tree, for which $\nu(R)=1$. Here $R$ is some
vertex of the tree ${\cal T}(X)$ (the root of the tree).

2) Monotonicity and infinitesimality: \be\label{mult}
\eta(\mu(J))>\eta(\nu(I)) \ee for $I>J$ and, moreover,
\be\label{mult1} d\eta(\nu(J))=\eta(\nu(J))-\eta(\nu({J+1}))
\qquad \hbox{is~ a~ positive~ infinitesimal~ value}\ee where $J+1$
is the smallest vertex larger than $J$.

3) Vanishing of the limit:
\be\label{dpdp}\lim_{I\to\infty}\eta(\nu(I))=0.\ee

\bigskip

We will perform computations with subtrees ${\cal S}\subset {\cal
T}$ of regular type. We will take $R\in {\cal S}_{\rm min}$ and
will claim, that $\eta(\nu(J))$ for $J\in {\cal S}_{\rm min}$
should be equal to 1 up to infinitesimal corrections which we will
neglect.

The rule (\ref{dpdp}) means that the limit $n\to 0$ is related to
the limit $I\to\infty$ in the directed tree. Thus our construction
indeed is a variant of the $n\to 0$ limit, since in our approach
$\nu(I)$ coincides with the dimension $n$ of the replica matrix
(when $I$ is the maximal vertex in the subtree ${\cal S}$ of the
regular type).

Condition (\ref{mult1}) may be discussed as the corollary of the
fact that in the $n\to0$ limit for any $J$ one has $p_J\to
1-\varepsilon$ for infinitesimal $\varepsilon$ (since the ball
${J+1}$, roughly speaking, is $p_{J+1}$ times larger than the ball
$J$). Some variant of the analogous construction was described in
paper \cite{PaSu}, where, in the $p$--adic case, the $n\to 0$
limit was discussed as the map $p\mapsto 1-\varepsilon$,
$\varepsilon\to 0$, which is the analogue of the formula
(\ref{mult1}).

Then, we introduce the $n\to 0$ limit in the RSB anzats under
consideration as the map $\rho$, which acts on the polynomials
over the variables, equal to the measures of the ultrametric disks
$\nu(J)$. This map is linear with respect to addition and
multiplication by numbers, and action on the monomials of $\nu(J)$
is defined as follows: \be\label{namonomy}
\rho\left(\nu^{k}(J)\right)=\nu(J)\eta^{k-1}\left(\nu(J)\right)
\ee

The formulas (\ref{norm42})--(\ref{dpdp}) are the direct analogues
of the definitions of the $n\to0$ limit for the Parisi anzats, and
the formula (\ref{namonomy}) is the new condition which was
trivial for the Parisi anzats, an becomes nontrivial in the case
under consideration.

\section{The $n\to 0$ limit: computation}

Investigate the introduced $n\to0$ limit construction for the
case, which is the direct generalization of the Parisi anzats for
the case of general ultrametric space.

\begin{lemma}{\sl For the case, when the replica matrix is defined
by the function of the measures of ultrametric disks \be\label{QP}
T{(J)}=F(\nu(J)) \ee the functional \be\label{thefunc} {1\over
\nu(K)}\sum_{J\in {\cal S}\backslash{\cal S}_{\rm
min}}T{(J)}\Delta \nu^2(J) \ee in the $n\to0$ limit, in the case
when the function $F$ is continuous on the interval $[0,1]$, takes
the form \be\label{limthefunc2} \lim_{n\to0}{1\over
n}\sum_{ab}Q_{ab}=-\int_0^1 F(x)dm(x) \ee where the integral on
the interval $[0,1]$ is defined as
$$
\int_0^1 F(x)dm(x)=\lim_{K\to\infty}{1\over \nu(K)}\sum_{J\in
{\cal S}}F(\rho(\nu({J})))\nu({J})d\rho(\nu({J}))
$$
i.e. the measure $dm(x)$ is defined as
$$
m([a,b])=\lim_{K\to\infty}{1\over \nu(K)}\sum_{J\in {\cal
S}:\rho(\nu(J))\in [a,b]}\nu({J})d\rho(\nu({J}))
$$
for $[a,b]\subset [0,1]$.
 }
\end{lemma}

\noindent{\it Proof}\qquad By (\ref{LeibRule}) we have
$$
\rho(\Delta\nu^2(J))=-\sum_{j=0}^{p_J-1}\nu({J_j})d\rho(\nu({J_j}))
$$
Therefore
$$
\rho\left[{1\over \nu(K)}\sum_{J\in {\cal S}\backslash{\cal
S}_{\rm min}}T{(J)}\Delta \nu^2(J)\right]=-{1\over
\nu(K)}\sum_{J\in {\cal S}\backslash
K}\rho\left[T{(J+1)}\right]\nu({J})d\rho(\nu({J}))
$$
For the case $T{(J)}=F(\nu(J))$ this takes the form
$$
-{1\over \nu(K)}\sum_{J\in {\cal S}\backslash
K}F(\rho(\nu({J}))-d\rho(\nu({J})))\nu({J})d\rho(\nu({J}))
$$
which for continuous $F$ reduces to
$$
-{1\over \nu(K)}\sum_{J\in {\cal S}\backslash
K}F(\rho(\nu({J})))\nu({J})d\rho(\nu({J}))
$$
Application of the $K\to\infty$ limit proves the lemma.

\bigskip

\noindent{\bf Remark} \qquad It is interesting to discuss this
result for the case when $T{(J)}=T={\rm const}$. Consider the
identity
$$
T{(J)}\Delta\nu^2(J)=T{(J)}\nu^2(J)-\sum_{j=0}^{p_J-1}T{(J_j)}\nu^2({J_j})
+\sum_{j=0}^{p_J-1}T{(J_j)}\nu^2({J_j})-T{(J)}\sum_{j=0}^{p_J-1}\nu^2({J_j})=
$$
$$
=\Delta\left[T{(J)}\nu^2(J)\right]-\sum_{j=0}^{p_J-1}\left[T{(J)}-T{(J_j)}\right]\nu^2({J_j})
$$
Applying the tree Newton--Leibnitz formula, we get for the
functional (\ref{thefunc})
$$
{1\over \nu(K)}\sum_{J\in {\cal S}\backslash {\cal S}_{\rm
min}}T{(J)}\Delta\nu^{2}(J)=
$$
$$
={1\over \nu(K)}\sum_{J\in {\cal S}\backslash {\cal S}_{\rm
min}}\left[\Delta\left[T{(J)}\nu^2(J)\right]-\sum_{j=0}^{p_J-1}\left[T{(J)}-T{(J_j)}\right]\nu^2({J_j})\right]=
$$
$$
=T{(K)}\nu(K)-{1\over \nu(K)}\sum_{J\in {\cal S}_{\rm
min}}T{(J)}\nu^2(J)-{1\over \nu(K)}\sum_{J\in {\cal S}\backslash
{\cal S}_{\rm
min}}\sum_{j=0}^{p_J-1}\left[T{(J)}-T{(J_j)}\right]\nu^2({J_j})
$$
which for the case $T({J})=T={\rm const}$ reduces to
$$
T\left(\nu(K)-{1\over \nu(K)}\sum_{J\in {\cal S}_{\rm
min}}\nu^2(J)\right)
$$
Then application of the $n\to0$ limit gives
$$
T\lim_{K\to\infty}\left(\rho(\nu(K))-{1\over \nu(K)}\sum_{J\in
{\cal S}_{\rm min}}\nu(J)\rho(\nu(J))\right)=-T
$$
We see that the computation of this value includes the summation
$\sum_{J\in {\cal S}_{\rm min}}\nu(J)$, i.e. the integration of
the measure $\nu$ over the ultrametric space.

\section{Replica symmetry breaking equation}

In the present section we, using variational procedure and the
introduced analysis on trees, find the equation, which describes
replica symmetry breaking for the Sherrington--Kirkpatrick model
in the vicinity of phase transition (in other words, when matrix
elements of the replica matrix can be considered as small
parameters). In this vicinity the free energy can be decomposed
into the series of the functionals (of the type of traces of the
degrees) of the replica matrix. Functionals of this kind
(corresponding to the first several terms of the series) we
computed in Section 4. Let us compute variations of these
functionals with respect to variations of the matrix elements
$T{(J)}$.

\begin{lemma}\label{DtraceRMst}{\sl
Variations of the following functionals have the form

\be\label{varQ2} \delta\,{\rm tr}\,Q^2=\delta\, \sum_{J\in {\cal
S}\backslash {\cal S}_{\rm min}}{T{(J)}}^2\Delta\nu^2(J)=
$$
$$
= \sum_{J\in {\cal S}\backslash {\cal S}_{\rm min}}2T{(J)}\delta
T{(J)}\Delta\nu^2(J) \ee \be\label{varQ21}
\delta\sum_{ij}Q^4_{ij}=\delta\sum_{J\in {\cal S}\backslash {\cal
S}_{\rm min}}{T{(J)}}^4\Delta\nu^2(J)=
$$
$$
=\sum_{J\in {\cal S}\backslash {\cal S}_{\rm
min}}4{T{(J)}}^3\delta T{(J)}\Delta\nu^2(J) \ee }
\end{lemma}

\begin{lemma}\label{DtrQ3}{\sl Variation of the cubic functional takes the form

\be\label{ultra} \delta\,\,{\rm tr}\,Q^3
=\delta\,\biggl[\sum_{B\in {\cal S}\backslash {\cal S}_{\rm
min}}\left(3\nu(B)\Delta \nu^2(B)-2\Delta
\nu^3(B)\right){T{(B)}}^3+
$$
$$
+3\sum_{B\in {\cal S}\backslash {\cal S}_{\rm
min}}\Delta\nu^2(B)\sum_{L:B<L\le
K}(\nu(L)-\nu({L-1,B})){T{(L)}}^2T{(B)} \biggr]=
$$
$$
=3\sum_{B\in {\cal S}\backslash {\cal S}_{\rm min}}\delta
T{(B)}\biggl[\left(3\nu(B)\Delta \nu^2(B)-2\Delta
\nu^3(B)\right){T{(B)}}^2+
$$
$$
+2{T{(B)}}\sum_{L\in {\cal S}\backslash {\cal S}_{\rm
min}:L<B}(\nu(B)-\nu({B-1,L}))T{(L)}\Delta\nu^2(L) +
$$
$$
+\Delta\nu^2(B)\sum_{L:B<L\le
K}(\nu(L)-\nu({L-1,B})){T{(L)}}^2\biggr]\ee }
\end{lemma}

\noindent{\it Proof}\qquad  We have
$$
\delta\,\,{\rm tr}\,Q^3 =\sum_{B\in {\cal S}\backslash {\cal
S}_{\rm min}}\left(3\nu(B)\Delta \nu^2(B)-2\Delta
\nu^3(B)\right)3{T{(B)}}^2\delta T{(B)}+
$$
$$
+3\sum_{B\in {\cal S}\backslash {\cal S}_{\rm
min}}\Delta\nu^2(B)\sum_{L:B<L\le
K}(\nu(L)-\nu({L-1,B}))\left(2{T{(L)}}T{(B)}\delta T{(L)}
+{T{(L)}}^2\delta T{(B)}\right) =
$$
$$
=\sum_{B\in {\cal S}\backslash {\cal S}_{\rm
min}}\left(3\nu(B)\Delta \nu^2(B)-2\Delta
\nu^3(B)\right)3{T{(B)}}^2\delta T{(B)}+
$$
$$
+6\sum_{L\in {\cal S}\backslash {\cal S}_{\rm
min}}T{(L)}\Delta\nu^2(L)\sum_{B:L<B\le
K}(\nu(B)-\nu({B-1,L})){T{(B)}}\delta T{(B)} +
$$
$$
+3\sum_{B\in {\cal S}\backslash {\cal S}_{\rm
min}}\Delta\nu^2(B)\delta T{(B)}\sum_{L:B<L\le
K}(\nu(L)-\nu({L-1,B})){T{(L)}}^2=
$$
$$
=\sum_{B\in {\cal S}\backslash {\cal S}_{\rm
min}}\left(3\nu(B)\Delta \nu^2(B)-2\Delta
\nu^3(B)\right)3{T{(B)}}^2\delta T{(B)}+
$$
$$
+6\sum_{B\in {\cal S}\backslash {\cal S}_{\rm min}}{T{(B)}}\delta
T{(B)}\sum_{L\in {\cal S}\backslash {\cal S}_{\rm
min}:L<B}(\nu(B)-\nu({B-1,L}))T{(L)}\Delta\nu^2(L) +
$$
$$
+3\sum_{B\in {\cal S}\backslash {\cal S}_{\rm
min}}\Delta\nu^2(B)\delta T{(B)}\sum_{L:B<L\le
K}(\nu(L)-\nu({L-1,B})){T{(L)}}^2=
$$
$$
=3\sum_{B\in {\cal S}\backslash {\cal S}_{\rm min}}\delta
T{(B)}\biggl[\left(3\nu(B)\Delta \nu^2(B)-2\Delta
\nu^3(B)\right){T{(B)}}^2+
$$
$$
+2{T{(B)}}\sum_{L\in {\cal S}\backslash {\cal S}_{\rm
min}:L<B}(\nu(B)-\nu({B-1,L}))T{(L)}\Delta\nu^2(L) +
$$
$$
+\Delta\nu^2(B)\sum_{L:B<L\le
K}(\nu(L)-\nu({L-1,B})){T{(L)}}^2\biggr]
$$
Here we used the transformation from lemma  \ref{change} (change
of order of integration in the tree integral). This finishes the
proof of the lemma.

\bigskip

Consider the functional which approximates the free energy of the
Sherrington--Kirkpatrick model near phase transition, when the
replica matrix can be considered as a small parameter. This
functional, which can be obtained by decomposition of the free
energy into the Taylor series and summation over the spin degrees
of freedom, has the form \cite{MPV} \be\label{theF} F=a_2{\rm
tr}\,Q^2+a_3{\rm tr}\,Q^3+a_4\sum_{ij}Q^4_{ij} \ee where $a_2$,
$a_3$, $a_4$ are some constants. To obtain the replica solution,
one has to vary this functional, in the framework of the replica
anzats under consideration, over the parameters of the anzats and
consider the equation
$$
\delta F=0
$$
which is called the replica symmetry breaking equation.

For the replica anzats under consideration we vary the free energy
over the parameters $T{(L)}$. Combining the lemmas
\ref{DtraceRMst}, \ref{DtrQ3}, we get the following theorem:

\begin{theorem}\label{varequ}{\sl Replica symmetry breaking equation $\delta F=0$
for the free energy (\ref{theF}) in the frameworks of replica
anzats (\ref{newmatrix}) takes the form \be\label{var_problem}
2a_2T{(B)}\Delta\nu^2(B)+4a_4{T{(B)}}^3\Delta\nu^2(B)+
$$
$$
+3a_3\biggl[\left(3\nu(B)\Delta \nu^2(B)-2\Delta
\nu^3(B)\right){T{(B)}}^2+
$$
$$
+2{T{(B)}}\sum_{L\in {\cal S}\backslash {\cal S}_{\rm
min}:L<B}(\nu(B)-\nu({B-1,L}))T{(L)}\Delta\nu^2(L) +
$$
$$
+\Delta\nu^2(B)\sum_{L:B<L\le
K}(\nu(L)-\nu({L-1,B})){T{(L)}}^2\biggr]=0 \ee }
\end{theorem}

\section{The constant RSB solution}

In the present section we check, that the constant replica matrix,
for which $T{(J)}=T={\rm const}$, indeed is a solution of
(\ref{var_problem}) (in the $n\to 0$ limit). If we substitute
$T{(J)}=T$ into (\ref{var_problem}), we get \be\label{checkRSym}
2a_2T\Delta\nu^2(B)+4a_4 T^3\Delta\nu^2(B)+
$$
$$
+3a_3T^2\biggl[3\nu(B)\Delta \nu^2(B)-2\Delta \nu^3(B)+
$$
$$
+2\sum_{j=1}^{p_B-1}(\nu(B)-\nu({B_j}))\left(\nu^2({B_j})-\sum_{L\in{\cal
S}_{\rm min}:L<B_j}\nu^2(L)\right) +
$$
$$
+\Delta\nu^2(B)(\nu(K)-\nu({B}))\biggr]=0\ee

This transforms into
$$
2a_2T\Delta\nu^2(B)+4a_4 T^3\Delta\nu^2(B)+
3a_3T^2\biggl[\nu(K)\Delta \nu^2(B)-
$$
$$
-2\sum_{j=1}^{p_B-1}(\nu(B)-\nu({B_j}))\sum_{L\in{\cal S}_{\rm
min}:L<B_j}\nu^2(L) \biggr]=0
$$
Apply the $n\to 0$ limit (i.e. the map $\rho$). We get, taking
into account that $\rho(\nu(K))\to0$ for $K\to\infty$ and
$\rho(\nu(L))\to1$ for $L\in{\cal S}_{\rm min}$
$$
2a_2T\rho(\Delta\nu^2(B))+4a_4 T^3\rho(\Delta\nu^2(B))-
3a_3T^2\biggl[ 2\sum_{j=1}^{p_B-1}\rho(\nu(B)-\nu({B_j}))
\nu({B_j})\biggr]=0
$$
Since the combination in square brackets is exactly $2\rho(\Delta
\nu^2(B))$, we get the equation
$$
\left(2a_2 T+4a_4 T^3 -6 a_3T^2\right)\rho(\Delta \nu^2(B))=0
$$
which has the solution $T=0$ (trivial), and the solutions
$$
T={3a_3\pm\sqrt{9a^2_3-8a_2a_4}\over 4a_4}=\mp{a_2\over 3a_3}
$$
The last equality holds if $a_2$ is a small parameter (which is
satisfied for the Sherrington--Kirkpatrick model in the considered
regime).

For the Parisi RSB anzats the obtained solution reduces to the
solution with unbroken replica symmetry. For the considered RSB
anzats the constant RSB solution depends on general ultrametric
space $X$ and a measure of generic form on this space, and
therefore it describes a wide class of physical situations.

\section{Appendix: Ultrametric PDO}

\begin{definition}{\sl
An ultrametric space is a metric space with the ultrametric $|xy|$
(where $|xy|$ is called the distance between $x$ and $y$), i.e.
the function of two variables, satisfying the properties of
positivity and non degeneracy
$$
|xy|\ge 0,\qquad |xy|=0\quad \Longrightarrow\quad x=y;
$$
symmetricity
$$
|xy|=|yx|;
$$
and the strong triangle inequality
$$
|xy|\le{\rm max }(|xz|,|yz|),\qquad \forall z.
$$
}
\end{definition}

Consider an ultrametric space $X$, satisfying the following
properties:

\medskip

1) The set of all the balls of nonzero diameter in $X$ is no more
than countable;

\medskip

2) For any two balls $I$, $J$ in $X$ and any sequence of balls
$\{D^{(k)}\}$, for which $I\subset D^{(k)}\subset J$ for all $k$,
the sequence $\{D^{(k)}\}$ must be finite;

\medskip

3) Any ball is a finite union of maximal subballs.

\bigskip

For ultrametric space $X$ consider the set ${\cal T}(X)$, which
contains all the balls in $X$ of nonzero diameter, and the balls
of zero diameter which are maximal subbals in balls of nonzero
diameter. On this tree there is the structure of directed set,
generated by the order on balls (ordered by inclusion).

If a non--maximal vertex $I\in {\cal T}$ is incident to $p_I+1$
edges, we will say that the branching index of $I$ is $p_I$. If
maximal index $I\in {\cal T}$ is incident to $p_I$ edges, we will
say that the branching index of $I$ is $p_I$. Equivalently,
branching index of a vertex $I$ in directed tree is the number of
maximal elements, which less than $I$. The following theorem can
be found in \cite{nextIzv} (the analogous result was obtained in
\cite{Lemin}, see also \cite{Freud} where it was presented).

\begin{theorem}\label{the4}
{\sl The set ${\cal T}(X)$ which corresponds to the ultrametric
space $X$, satisfying properties (1), (2), (3) above with the
partial order, defined by inclusion of balls, is a directed tree
where all neighbor vertices are comparable.

Branching index for vertices of this tree may take finite integer
non--negative values not equal to one, and the maximal vertex (if
exists) has the branching index $\ge 2$. Balls of nonzero diameter
in $X$ correspond to vertices of branching index $\ge 2$ in ${\cal
T}(X)$, and the balls of zero diameter which are maximal subbals
in balls of nonzero diameter correspond to vertices of branching
index 0 in ${\cal T}(X)$. }
\end{theorem}

Remind that a directed set is a partially ordered set, where for
any pair of elements there exists the unique supremum with respect
to the partial order.

Consider the set $X\bigcup {\cal T}(X)$, where we identify the
balls of zero diameter from ${\cal T}(X)$ with the corresponding
points in $X$. We call ${\cal T}(X)$ the tree of balls in $X$, and
$X\bigcup {\cal T}(X)$ the extended tree of balls. One can say
that $X\bigcup {\cal T}(X)$ is the set of all the balls in $X$, of
nonzero and zero diameter.

Introduce the structure of a directed set on $X\bigcup {\cal
T}(X)$. At the tree ${\cal T}(X)$ this structure is the following:
$I<J$ if for the corresponding balls $I\subset J$.

The supremum
$$
{\rm sup}(x,y)=I
$$
of points $x,y\in X$ is the minimal ball $I$, containing the both
points.

Analogously, for $J\in {\cal T}(X)$ and $x\in X$ the supremum
$$
{\rm sup}(x,J)=I
$$
is the minimal ball $I$, which contains the ball $J$ and the point
$x$.

Conversely, starting from a directed tree one can reproduce the
corresponding ultrametric space \cite{Izv}, \cite{nextIzv},
\cite{Lemin}, \cite{Serre}, which will be the absolute of the
directed tree.

Consider a measure $\nu$ on ultrametric space $X$, such that for
arbitrary ball $D$ its measure $\nu(D)$ is a positive number (i.e.
is not equal to zero or to infinity).

We study the ultrametric pseudodifferential operator (or the PDO)
of the form considered in \cite{Izv}, \cite{nextIzv}
$$
Tf(x)=\int T{({\rm sup}(x,y))}(f(x)-f(y))d\nu(y)
$$
Here $T{(I)}$ is some function on the tree ${\cal T}( X)$. Thus
the structure of this operator is determined by the direction on
$X\bigcup {\cal T}(X)$. This kind of ultrametric PDO we call {\it
the {\rm sup}--operator}.

\bigskip

\centerline{\bf Acknowledgements}

\medskip

The authors would like to thank G.Parisi and I.V.Volovich  for
fruitful discussions and valuable comments.  One of the authors
(A.Kh.) would like to thank S.Albeverio for fruitful discussions
and support of $p$--adic investigations. This paper has been
partly supported by EU-Network ''Quantum Probability and
Applications''. One of the authors (S.V.Kozyrev) has been partly
supported by the CRDF (grant UM1--2421--KV--02), by The Russian
Foundation for Basic Research (project 05-01-00884-a), by the
grant of the President of Russian Federation for the support of
scientific schools NSh 1542.2003.1, by the Program of the
Department of Mathematics of Russian Academy of Science ''Modern
problems of theoretical mathematics'', by the grant of The Swedish
Royal Academy of Sciences on collaboration with scientists of
former Soviet Union, and by the grants DFG Project 436 RUS
113/809/0-1 and RFFI 05-01-04002-NNIO-a.

\end{document}